\documentclass[]{JHEP3} 

\usepackage{epsfig,multicol}
\usepackage{graphicx}
\usepackage{dcolumn}
\usepackage{bm}

\newcommand{\Tlss}{T_{\rm D}}
\newcommand{\alphalss}{\alpha_{\rm D}}
\newcommand{\etalss}{\eta_{\rm D}}

\title{CMB observations in LTB universes: Part I\\
--- Matching peak positions in the CMB spectrum ---
}

\author{Chul-Moon Yoo
\\
	Yukawa Institute for Theoretical Physics, Kyoto University
Kyoto 606-8502, Japan
\\
Asia Pacific Center for Theoretical
Physics, Pohang, Gyeongbuk 790-784, Republic of Korea
	E-mail: \email{yoo@yukawa.kyoto-u.ac.jp}}

\author{Ken-ichi Nakao
\\
Department of Mathematics and Physics,
Graduate School of Science, Osaka City University,
3-3-138 Sugimoto, Sumiyoshi, Osaka 558-8585, Japan
	E-mail: \email{knakao@sci.osaka-cu.ac.jp}}

\author{Misao Sasaki
\\
	Yukawa Institute for Theoretical Physics, Kyoto University
Kyoto 606-8502, Japan
\\
Korea Institute for Advanced Study
207-43 Cheongnyangni 2-dong, Dongdaemun-gu, 
Seoul 130-722, Republic of Korea
	E-mail: \email{misao@yukawa.kyoto-u.ac.jp}}

\received{\today} 		
\accepted{\today}		

\preprint{YITP-10-34,APCTP-Pre2010-003,OCU-PHYS-331,AP-GR-77}

\abstract{Acoustic peaks in the spectrum of the cosmic microwave background in 
spherically symmetric inhomogeneous cosmological models are studied. 
At the photon-baryon decoupling epoch, the universe may be assumed
to be dominated by non-relativistic matter
, 
and thus we may treat radiation as a test field in the 
universe filled with dust which is described by the Lema\^itre-Tolman-Bondi 
(LTB) solution. First, we give an LTB model 
whose distance-redshift relation agrees with 
that of the concordance $\Lambda$CDM model in the whole redshift domain 
and which is well approximated by 
the Einstein-de Sitter universe at and before decoupling. 
We determine the decoupling epoch in this LTB universe 
by Gamow's criterion and then calculate the positions of acoustic peaks. 
Thus obtained results are not consistent with the WMAP data. 
However, we find that one can fit the peak positions
by appropriately modifying the LTB model,
namely, by allowing the deviation of the distance-redshift relation
from that of the concordance $\Lambda$CDM model at $z>2$ where
no observational data are available at present.
Thus there is still a possibility of explaining the 
apparent accelerated expansion of the universe by inhomogeneity 
without resorting to dark energy if we abandon the Copernican principle.
Even if we do not take this extreme attitude,
it also suggests that local, isotropic inhomogeneities around us
may seriously affect the determination of the density contents
of the universe unless the possible existence of
such inhomogeneities is properly taken into account.
}


\begin{document} 


\section{Introduction}
\label{sec:intro}

In the study of cosmology, we usually assume that we do not live 
in a special position in the universe, the so-called Copernican principle.  
However, 
although this principle is 
natural, 
it should not be blindly assumed but should be justified observationally.
Conventionally observational data are interpreted under the assumption
of a homogeneous and isotropic universe on average. Therefore it is
not clear at all how big the systematic errors would be in the determination
of the cosmological parameters if this assumption were abandoned.
In other words, it is important to investigate possible
``anti-Copernican'' models of the universe and test if such models can
be observationally excluded. In this paper, as one of such attempts,
we try to construct a cosmological model without dark energy which is
consistent with the observed distance-redshift relation as well as
with the WMAP data.

In anti-Copernican models, we are assumed to be located at a 
special place in the universe, usually at the center of a 
spherically symmetric inhomogeneous universe~\cite{Zehavi:1998gz,
Tomita:1999qn,Tomita:2000jj,Tomita:2001gh,Celerier:1999hp}. 
In recent years, such models have attracted much 
attention~\cite{Alnes:2005rw,Alnes:2006pf,Alnes:2006uk,
Alexander:2007xx,Bolejko:2008cm,Enqvist:2007vb,Enqvist:2006cg,
February:2009pv,GarciaBellido:2008nz,GarciaBellido:2008yq,
GarciaBellido:2008gd,Garfinkle:2006sb,Kasai:2007fn,
2000ApJ...529...26T,Tomita:2000rf,Zibin:2008vk,Regis:2010iq,Clifton:2009kx,
Kodama:2010gr}, 
and various ways to observationally test these models
have been proposed by many authors~\cite{Biswas:2007gi,
Bolejko:2005fp,Bolejko:2008xh,Bolejko:2008ya,
Brouzakis:2006dj,Brouzakis:2007zi,Bhattacharya:2009bz,Clarkson:2007pz,
Caldwell:2007yu,Clifton:2008hv,Dabrowski:1997sm,Goodman:1995dt,
Godlowski:2004gh,Jia:2008ti,Lasky:2010vn,Marra:2007pm,Moffat:2006ct,
PascualSanchez:1999zr,Quartin:2009xr,Romano:2007zz,Stelmach:2006zc,
Tanimoto:2009mz,Tomita:2009yx,Uzan:2008qp}. 

One of the simplest ways to construct an anti-Copernican model 
is to solve the inverse problem of the distance-redshift relation. 
Although the isotropy of the universe around us has been confirmed
with high accuracy by the observation of the cosmic microwave 
background (CMB), this does not automatically imply
homogeneity of the universe.
Thus, in solving the inverse problem, we may assume
that the universe is spherically symmetric around us.
In addition, we usually assume that the universe is
dominated by cold dark matter, that is, by dust. 
The spherically symmetric dust filled spacetime is described by the 
Lema\^itre-Tolman-Bondi (LTB) solution. The LTB solution
has three arbitrary functions of the radial coordinate
approximately corresponding to the density profile, 
the spatial curvature and the big-bang time perturbation,
with one of them being a gauge degree of freedom representing 
the choice of the radial coordinate.
These arbitrary functions may be determined by requiring that
the resulting LTB universe be consistent with selected important 
observational data (e.g., the distance-redshift relation). 
However, we should note that it is not apparent at all if these three 
functions have enough degrees of freedom to fit all of the important
observational data.

In 1999, C\'{e}l\'{e}rier solved the inverse problem 
analytically at small redshifts $z\ll1$
in the form of the Maclaurin series~\cite{Celerier:1999hp}. 
Then, in 2002, Iguchi, Nakamura and Nakao constructed numerically 
an LTB model whose distance redshift relation agrees
with that of the $\Lambda$CDM model at $z\lesssim1.6$~\cite{Iguchi:2001sq}
\footnote{Before Refs.~\cite{Celerier:1999hp,Iguchi:2001sq}, Mustapha et.al 
discussed an inverse problem using the LTB solution~\cite{Mustapha:1998jb}.  
}. 
However, they could not go beyond $z\sim1.6$ due to a technical problem. 
Since then the inverse problem has been discussed 
by various authors~\cite{Chung:2006xh,Vanderveld:2006rb,Celerier:2009sv,
Kolb:2009hn,Romano:2009mr,Romano:2009ej,Yoo:2008su}.
In 2008, Yoo, Kai and Nakao succeeded in 
constructing an LTB model whose distance-redshift relation
agrees with that of the concordance $\Lambda$CDM model 
in the whole redshift domain and which is homogeneous 
in the early stage of the universe, that is, the big-bang
time perturbation being taken to be zero~\cite{Yoo:2008su}. 
Note that because of the existence of a gauge degree of
freedom in the choice of the radial coordinate, there remains
only one functional degree of freedom. In \cite{Yoo:2008su},
this is represented by the function that determines
the spatial curvature.

Not only the distance-redshift relation but also 
CMB observations in inhomogeneous cosmology
have been widely 
discussed~\cite{Alnes:2006pf,Alnes:2005rw,Alexander:2007xx,
Bolejko:2008cm,Clarkson:1999yj,Caldwell:2007yu,GarciaBellido:2008gd,
GarciaBellido:2008nz,GarciaBellido:2008yq,Goodman:1995dt,
Godlowski:2004gh,Zibin:2008vj,Zibin:2008vk,Regis:2010iq,Clifton:2009kx,Kodama:2010gr}. 
In many of these works, it is assumed that the universe 
is homogeneous in the spacelike asymptotic region 
from which the CMB photons come, hence
the CMB photon distribution is assumed to
be the same as that in the homogeneous and isotropic 
universe models at the time of decoupling. 
Several authors proposed parametrized LTB models and gave 
constraints on the parameters 
with observational data~\cite{Alnes:2005rw,Alexander:2007xx,
Bolejko:2008cm,Enqvist:2007vb,GarciaBellido:2008nz,
GarciaBellido:2008yq,GarciaBellido:2008gd,Zibin:2008vk}.

Recently, Bolejko and Wyithe~\cite{Bolejko:2008cm} suggested that 
it is possible to construct an LTB model which is consistent 
not only with the distance-redshift relation but also with
the acoustic peak positions of the WMAP data~\cite{Hinshaw:2006ia} 
by choosing appropriately the arbitrary functions of the LTB solution. 
In this paper, 
by modifying the LTB model given in Ref.~\cite{Yoo:2008su}
we show explicitly that this is indeed possible. 
For simplicity, we use Gamow's criterion to determine
the decoupling epoch instead of invoking a precise 
numerical calculation. 
This is the same procedure adopted in Ref.\cite{Regis:2010iq}. 
An advantage of this simplified prescription is that
it makes it easy to understand the physical degrees of freedom 
in the LTB universe that determine the CMB anisotropy spectrum.
We note that, in our construction, only the asymptotic homogeneity 
on the past light cone of the observer at the symmetry center 
was necessary in contrast to a much stronger
asymptotic spatial homogeneity condition
(see a related discussion in Ref.~\cite{Enqvist:2009hn}).

This paper is organized as follows. 
In \S\ref{sec:background}, we give a brief review
of the LTB solution and 
give a 
fitting function for the spatial curvature
of the LTB model that has the exactly same distance-redshift 
relation as the concordance $\Lambda$CDM model~\cite{Yoo:2008su}. 
In \S\ref{sec:LSSinLTB},
we discuss Gamow's criterion and 
the physical degrees of freedom at decoupling.
The position of the first acoustic peak in the CMB spectrum 
in an LTB universe is discussed in \S\ref{sec:comparison}. 
In \S\ref{sec:inverseproblem}, 
we construct an LTB model which is consistent with 
the acoustic peak positions of the WMAP data
by modifying the asymptotic structure of the LTB model
obtained in~\cite{Yoo:2008su}.  
\S\ref{sec:sudi} is devoted to summary and discussion. 

\section{LTB model from the inverse problem}
\label{sec:background}

As mentioned in the introduction,
we consider a spherically symmetric 
inhomogeneous universe filled with dust. 
This universe is described by an exact solution of 
the Einstein equations, known as the 
Lema\^itre-Tolman-Bondi (LTB) solution. 
The metric of the LTB solution is given by
\begin{equation}
ds^2=-c^2dt^2+\frac{\left(\partial_r R(t,r)\right)^2}{1-k(r)r^2}dr^2
+R^2(t,r)d\Omega^2, \label{eq:metric}
\end{equation}
where $k(r)$ is an arbitrary function of the radial coordinate $r$. 
The matter is dust whose stress-energy tensor is given by
\begin{equation}
T^{\mu\nu}=\rho u^\mu u^\nu,
\end{equation}
where $\rho=\rho(t,r)$ is the 
mass density, 
and $u^a$ is the four-velocity of 
the fluid element. 
The coordinate system in Eq.~(\ref{eq:metric}) is chosen
in such a way that $u^\mu=(1,0,0,0)$.

The area radius $R(t,r)$ 
satisfies one of the Einstein equations, 
\begin{equation}
\left(\frac{\partial R}{\partial t}\right)^2=\frac{2M(r)}{R}-k(r)r^2,
\label{eq:Einstein-eq}
\end{equation}
where $M(r)$ is an arbitrary function related to 
the 
mass 
density $\rho$ by
\begin{equation}
\rho(t,r)=\frac{1}{4\pi R^2(t,r)}\frac{dM(r)}{dr}.
\end{equation}
Following Ref.~\cite{Tanimoto:2007dq}, we write the solution 
of Eq.~(\ref{eq:Einstein-eq}) in the form,
\begin{eqnarray}
R(t,r)&=&(6GM(r))^{1/3}(t-t_{\rm B}(r))^{2/3} S(x), 
\label{eq:YS}\\
x&=&c^2k(r)r^2\left(\frac{t-t_{\rm B}(r)}{6GM(r)}\right)^{2/3}, \label{eq:defx}
\end{eqnarray}
where $t_{\rm B}(r)$ is an arbitrary function 
which determines the big bang time, 
and $S(x)$ is a function defined implicitly as
\begin{equation}
S(x)=
\left\{\begin{array}{lll}
\displaystyle
\frac{\cosh\sqrt{-\eta}-1}{6^{1/3}(\sinh\sqrt{-\eta}
-\sqrt{-\eta})^{2/3}}
\,;\qquad
&\displaystyle
x=\frac{-(\sinh\sqrt{-\eta}-\sqrt{-\eta})^{2/3}}{6^{2/3}}
\quad&\mbox{for}~~x<0\,,
\\
\displaystyle
\frac{1-\cos\sqrt{\eta}}{6^{1/3}(\sqrt{\eta}
-\sin\sqrt{\eta})^{2/3}}
\,;&\displaystyle
x=\frac{(\sqrt{\eta}-\sin\sqrt{\eta})^{2/3}}{6^{2/3}}
\quad&\mbox{for}~~x>0\,,
\end{array}\right.
\label{eq:defS}
\end{equation}
and $S(0)=({3}/{4})^{1/3}$. 
The function $S(x)$ is analytic for $x<(\pi/3)^{2/3}$. 
Some characteristics of the function $S(x)$ 
are given in Refs.~\cite{Yoo:2008su} and \cite{Tanimoto:2007dq}. 

As shown in the above, the LTB solution has three arbitrary functions, 
$k(r)$, $M(r)$ and $t_{\rm B}(r)$. 
One of them is a gauge degree of freedom 
for rescaling of the radial coordinate $r$.
In this paper, we fix this by setting
\begin{equation}
M(r)=\frac{4}{3}\pi\rho_0r^3, 
\end{equation}
where $\rho_0$ is the energy density at the 
symmetry center at present $\rho_0=\rho(t_0,0)$.
As in the case of the homogeneous and isotropic 
universe, the present Hubble parameter $H_0$ is related to $\rho_0$ as 
\begin{equation}
H_0^2+k(0)c^2=\frac{8}{3}\pi G\rho_0. 
\end{equation}
As in Ref.~\cite{Yoo:2008su}, we assume the simultaneous big bang, i.e., 
\begin{equation}
t_{\rm B}(r)=0. \label{eq:tB}
\end{equation}
For notational simplicity, we introduce 
dimensionless quantities,
\begin{eqnarray*}
\tilde r:=\frac{H_0r}{c}\,,\quad \tilde k(\tilde r):=\frac{k(r)c^2}{H_0^2}\,. 
\end{eqnarray*}

\begin{figure}[htbp]
\begin{center}
\includegraphics[scale=1.8]{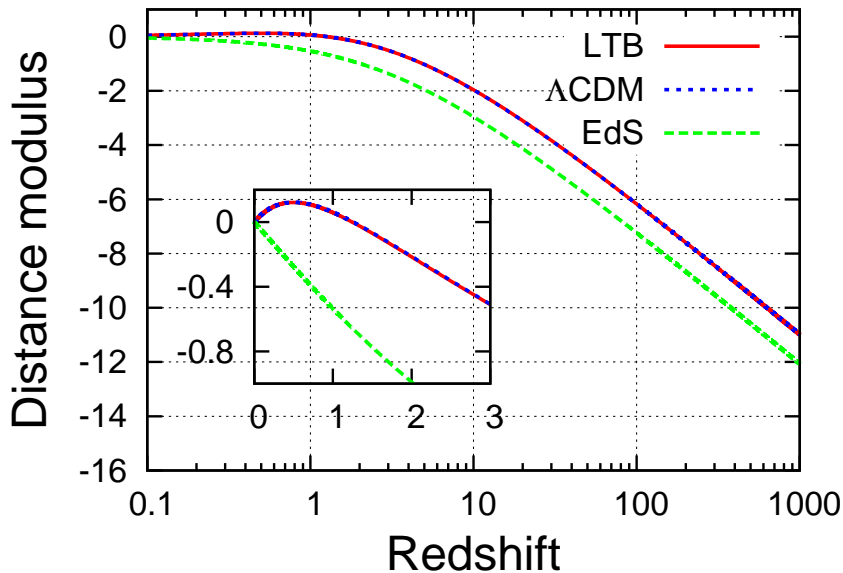}
\caption{Distance modulus for our LTB model and 
the concordance $\Lambda$CDM model. 
}
\label{fig:distance}
\end{center}
\end{figure}

The observed distance-redshift relation is 
consistent with the homogeneous and isotropic 
universe model with $(\Omega_{\rm m0},\Omega_{\Lambda0})=(0.3,0.7)$, 
where $\Omega_{\rm m0}$ is the density parameter of the
total non-relativistic matter (i.e., cold dark matter plus baryons)
and $\Omega_{\Lambda0}$ is that of the cosmological constant, 
the so-called concordance $\Lambda$CDM model.
We determine $\tilde k(\tilde r)$ so that the 
distance-redshift relation of our LTB model agrees with that of the 
concordance $\Lambda$CDM model. 
In Ref.~\cite{Yoo:2008su}, 
the inverse problem was solved numerically for $\tilde{k}(\tilde{r})$. 
A fitting function to the numerical result obtained in Ref.~\cite{Yoo:2008su}
is given by\footnote{It should be noted that the fitting function 
has a non-vanishing first derivative at the center, corresponding
to the existence of a spike in the density profile. 
In our model the central region has a void-like structure
with $\Omega_{\rm m0}\sim0.09$ at the center. 
However, the density quickly rises to $\Omega_{\rm m0}\sim 0.2$ 
at $z=0.3$ in agreement with large scale structure 
observations, as shown in Ref.~\cite{Yoo:2008su}.
The longitudinal Hubble parameter $H_L:=\partial_t\partial_rR/\partial_rR$ 
in our LTB model is about 10\% 
larger than that in the concordance $\Lambda$CDM model at $z\sim 2$. 
Although we need more careful investigations, 
considering large error bars and uncertainties 
in the local Hubble measurements~\cite{Simon:2004tf}, 
there seems no apparent conflict between our model and
observation.
}
\begin{equation}
\tilde k_{\rm fit}(\tilde r)
=\frac{0.545745}{0.211472+ \sqrt{0.026176+ \tilde r}} 
- \frac{2.22881}{\left(0.807782+ \sqrt{0.026176+ \tilde r}\right)^2}. 
\end{equation}
As shown in Fig.\ref{fig:distance}, the distance in our LTB universe model 
with $\tilde k(\tilde r)=\tilde k_{\rm fit}(\tilde r)$ agrees 
with that in the concordance $\Lambda$CDM model 
in the whole redshift domain.

\section{Decoupling Epoch in the LTB universe}
\label{sec:LSSinLTB}
The decoupling between photons and baryons occurs 
in an inhomogeneous universe just as in the case of
a homogeneous universe. That is, it occurs when the mean free path 
of photons becomes effectively infinite due to almost
complete recombination of electrons to protons.

Since there is no radiation component in our LTB model,
we cannot treat the decoupling in a rigorous manner.
However, as in the concordance model, we expect
the energy density of the radiation to be only a small 
fraction of the total density at decoupling, hence its effect
on the spacetime geometry is small, if not negligible. 
In fact, the radiation energy density estimated in our LTB model
turns out to be about 20\% of the total energy density.
This means that treating the radiation as a test field in
our model is consistent to a first approximation.

\begin{figure}[htbp]
\begin{center}
\includegraphics[scale=0.7]{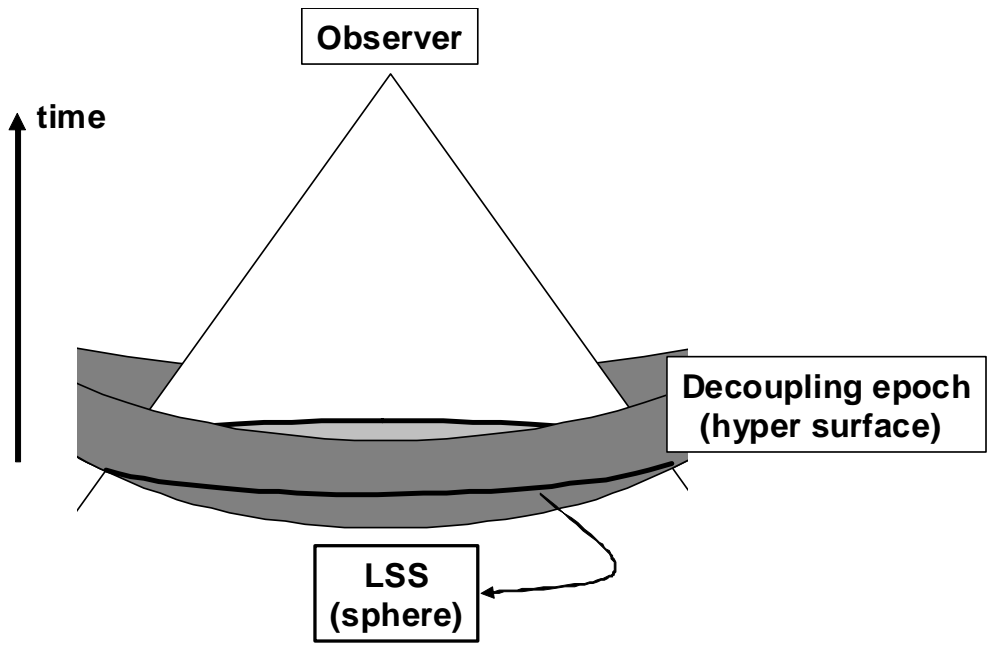}
\caption{Schematic figure for the hypersurface of the decoupling epoch 
in an LTB universe. 
}
\label{fig:lightconeandlss}
\end{center}
\end{figure}

Another approximation we adopt is the instantaneous decoupling
Namely, we assume decoupling to occur on a single spacelike
hypersurface. 
Since our LTB universe model is inhomogeneous but spherically symmetric,
it is natural to assume that the decoupling hypersurface is also 
inohomogeneous but spherically symmetric.
Thus it is specified by the form,
\begin{equation}
t=t_{\rm D}(r)\,.
\end{equation}
The cross section of this hypersurface with
the past directed null cone from the observer at the center
constitutes the last scattering surface (LSS) of CMB photons
 (see Fig.~\ref{fig:lightconeandlss}).
The LSS is a spacelike 2-dimensional sphere 
by the assumed symmetry.  
We use the subscript $*$ to express physical quantities 
on the LSS. In our approximation, ignoring secondary effects, 
the CMB anisotropy is essentially determined by the distribution
of photons on the LSS. 

The geodesic equations to determine the past light cone from
the observer at the center are written in the form,
\begin{eqnarray}
(1+z)\frac{dt}{dz}
&=&-\frac{\partial _r R}{\partial_t\partial_rR},
\label{eq:nullgeo1}\\
(1+z)\frac{dr}{dz}
&=&\frac{c\sqrt{1-k(r)r^2}}{\partial_t\partial_rR}, 
\label{eq:nullgeo2}
\end{eqnarray}
where the past directed radial null geodesics have been 
parametrized by the cosmological redshift $z$.
We denote the solution of the above equations by
\begin{equation}
t=t_{\rm lc}(z)~,~r=r_{\rm lc}(z). 
\end{equation}

Now to discuss the decoupling condition, we consider
the matter contents of the universe around the time of decoupling.
For simplicity, we consider a universe consists of
cold dark matter, protons and electrons, and neutral hydrogen atoms.
In particular, we neglect helium. Since the contribution of
these other components is not large, this simplification should
not lead to a serious error in our analysis.
We also assume that, until the decoupling time,
photons, electrons and protons are in thermal equilibrium. 
The energy density of the electrons and photons is negligible, and hence 
the constituents of our LTB model are cold dark dark matter 
and baryons, the latter of which consist of protons and hydrogen atoms.
Thus the baryon number density 
$n_{\rm b}$ is equal to the total number density of 
protons and hydrogen atoms, 
and the electron number density $n_{\rm e}$ is equal to
the proton number density.

In homogeneous and isotropic cosmology, 
the decoupling time is well determined by Gamow's criterion,
\begin{equation}
H= \Gamma,
\label{eq:Gamow1}
\end{equation}
where $H$ is the Hubble parameter and $\Gamma$ is the rate
of collisions of a photon with electrons.
Using the Thomson scattering cross section $\sigma_{\rm T}$ and the 
electron number density $n_{\rm e}$, $\Gamma$ is written as
\begin{equation}
\Gamma=cn_{\rm e}\sigma_{\rm T}\,.
\end{equation}
In our LTB model, we also adopt this Gamow's criterion,
with the identification of the``Hubble parameter $H$'' 
with 
\begin{equation}
H^2=\frac{8\pi G}{3}\rho\,.
\end{equation}
Since, by virtue of Eq.~(\ref{eq:tB}), our LTB universe model 
will be very similar to the Einstein-de Sitter universe near the 
decoupling time, the above definition should be accurate enough
for the present purpose. 

In order to estimate the electron number density, 
let us consider the ionization rate $X_{\rm e}:=n_{\rm e}/n_{\rm b}$. 
In the thermal equilibrium, the ionization rate $X_{\rm e}$ satisfies 
Saha's equation,
\begin{equation}
\frac{1-X_{\rm e}}{X_{\rm e}^2}=\frac{4\sqrt{2}\zeta(3)}{\sqrt{\pi}}\eta
\left(\frac{k_{\rm B}T}{m_{\rm e}c^2}\right)^{3/2}
\exp\left(\frac{13.59 {\rm eV}}{k_{\rm B}T}\right), 
\end{equation}
where $\zeta(x)$ is the zeta function, and 
$T$, $\eta$, $k_{\rm B}$ and $m_{\rm e}$ are the temperature, 
the baryon-to-photon ratio, 
the Boltzmann constant and the electron mass, respectively. 
Since the ionization rate at decoupling drops down to
$X_e\sim 10^{-5}$, we approximate the above equation
by 
\begin{equation}
X_{\rm e}^2\simeq\frac{\sqrt{\pi}}{4\sqrt{2}\zeta(3)}\frac{1}{\eta}
\left(\frac{k_{\rm B}T}{m_{\rm e}c^2}\right)^{-3/2}
\exp\left(-\frac{13.59{\rm eV}}{k_{\rm B}T}\right). 
\end{equation}
Using this equation, Gamow's criterion (\ref{eq:Gamow1}) 
is rewritten in the form,
\begin{equation}
\eta=\frac{32\sqrt{2\pi}\zeta(3)}{3} 
\frac{G\rho}{(cn_{\gamma 0}\sigma_{\rm T})^2}
\left(\frac{k_{\rm B}T_0}{m_{\rm e}c^2}\right)^{3/2}
\left(\frac{T_0}{T}\right)^{9/2}
\exp\left(\frac{13.59{\rm eV}}{k_{\rm B}T}\right), 
\label{eq:Gamow2}
\end{equation}
where $n_{\gamma0}$ and $T_0\simeq 2.725$K are the present 
photon number density and observed CMB temperature, respectively. 

Since we assume thermal equilibrium of electrons, 
protons and photons until the decoupling time $t=t_{\rm D}(r)$, 
the physical state of the decoupling hypersurface,
which is spherically symmetric, is determined by the 
distributions of the temperature $T=\Tlss(r)$, 
the baryon-to-photon ratio  $\eta=\etalss(r)$,
and the matter energy density $\rho=\rho(t_{\rm D}(r),r)$.
For convenience, in place of $\rho(t_{\rm D}(r),r)$,
we introduce the following quantity:
\begin{equation}
\alphalss(r):=\frac{\rho(t_{\rm D}(r),r)}{\rho_0}
\left(\frac{T_0}{\Tlss(r)}\right)^3. \label{eq:alpha}
\end{equation}
The quantity $\alphalss(r)$ is proportional 
to the ratio of the matter density 
and the photon number density. Then, from Eq.~(\ref{eq:Gamow2}), we obtain 
\begin{equation}
\etalss(r)=\frac{32\sqrt{2\pi}\zeta(3)}{3} 
\frac{G\rho_0}{(cn_{\gamma 0}\sigma_{\rm T})^2}
\left(\frac{k_{\rm B}T_0}{m_{\rm e}c^2}\right)^{3/2}
\alphalss(r)
\left(\frac{T_0}{\Tlss(r)}\right)^{3/2}
\exp\left(\frac{13.59{\rm eV}}{k_{\rm B}\Tlss(r)}\right).
\label{eq:gamow}
\end{equation}
We note that once $\Tlss(r)$ and $\alphalss(r)$ are given, 
the hypersurface $t=t_{\rm D}(r)$ can be obtained 
from Eq.~(\ref{eq:alpha}).

On the LSS, we must have
\begin{equation}
\frac{\Tlss(r_{\rm lc}(z_*))}{T_0}
=\frac{T_*}{T_0}=1+z_*. 
\label{eq:zlss}
\end{equation}
Then, if we regard $\Tlss(r)$, $\etalss(r)$ and $\alphalss(r)$ 
as mutually independent functions, 
we have one 
functional condition (\ref{eq:gamow}) and 
one boundary condition at $z=z_*$. 
to 
constrain these three functions. 
But these are not enough to determine the decoupling hypersurface, 
$t=t_{\rm D}(r)$, through Eq.~(\ref{eq:alpha}). 
We would need two more functional conditions.
However, for the present purpose, we do not need to determine the 
total shape of the decoupling hypersurface, 
but we only have to know the location of the LSS. 
Therefore we only need a single condition on the LSS 
in addition to Eqs.~(\ref{eq:gamow}) and (\ref{eq:zlss}).
In this paper, for simplicity, we impose 
\begin{equation}
\eta_*=6.2\times10^{-10}\,. \label{eq:eta-star}
\end{equation}

We numerically solve the radial null geodesic 
equations (\ref{eq:nullgeo1}) and (\ref{eq:nullgeo2}). 
At each redshift $z$, we define $T$ and $\alpha$ by  
\begin{eqnarray}
T&=&(1+z)T_0,\\
\alpha&=&\frac{\rho(t_{\rm lc}(z),r_{\rm lc}(z))}{\rho_0(1+z)^3}. 
\end{eqnarray}
During the numerical integration,
we check at each redshift $z$ whether Eq.~(\ref{eq:gamow}) 
is satisfied by $\Tlss=T$, $\alphalss=\alpha$ 
and $\etalss=\eta_*$. 
If Eq.~(\ref{eq:gamow}) is satisfied, we 
stop integrating Eqs.~(\ref{eq:nullgeo1}) and (\ref{eq:nullgeo2}), 
and identify $z$, $T$ and $\alpha$ 
at this moment with $z_*$, $T_*$ and $\alpha_*$, 
respectively. 
The result we have obtained is
\begin{eqnarray}
\frac{T_*}{T_0}&=&1+z_*\simeq1129\,, \label{eq:T-result}\\
\alpha_*&\simeq&4.856\,, \label{eq:alpha-result}
\end{eqnarray}
for our LTB universe model. 

In the above discussion, we have focused on only $\alpha_*$ and 
$\eta_*$, 
and we have given no further constraint on $\alphalss(r)$ and $\etalss(r)$. 
Therefore, we can assume $\alphalss(r)=\alpha_*$ and $\etalss(r)=\eta_*$ 
over the whole decoupling hypersurface. 
Inhomogeneities of $\alphalss$ and $\etalss$ correspond to an isocurvature 
perturbation of the matter density and a baryon isocurvature perturbation,
respectively.  
However this assumption 
might lead to a contradiction to 
the observational data of the kinematic Sunyaev-Zel'dovich effects 
as will be discussed in a forthcoming paper.
It is also worthy to notice that the 
anomalous primordial Lithium
abundances reported in Ref.\cite{Cyburt:2008kw} can be explained by 
introducing the baryon isocurvature
perturbation \cite{Regis:2010iq}. 

\section{acoustic peaks in the CMB spectrum}
\label{sec:comparison}

\subsection{acoustic peaks in the homogeneous and isotropic background}

Let us first briefly review the positions of the acoustic peaks 
in the CMB spectrum in the homogeneous 
and isotropic background universe.
The acoustic peak positions in the CMB spectrum 
can be written as \cite{Hu:2000ti}
\begin{equation}
\ell_m=(m-\phi_m)\pi\frac{d_{\rm A}}{r_{\rm s*}}, 
\label{eq:fitting}
\end{equation}
where $d_{\rm A}$ is the angular diameter distance from the LSS to
the observer at present, $r_{\rm s*}$ is the radius of the sound
horizon at the time of decoupling, and $\phi_m$ is a small
correction to the position of the $m$-th peak. 
As explicitly shown in the Appendix \ref{sec:peakposihome}, 
$r_{\rm s*}$ and $\phi_m$ (for $m=1,2,3$) can be expressed 
in terms of 
$\omega_{\rm m}:=\Omega_{\rm m0}h^2$, $\omega_{\rm b}:=\Omega_{\rm b0}h^2$, 
$\omega_\gamma:=\Omega_{\gamma0}h^2$, 
$z_*$ and the spectral index $n_{\rm s}$ 
which characterizes the spectrum of the initial 
density perturbation, 
where $\Omega_{\rm m0}$, $\Omega_{\rm b0}$, $\Omega_{\rm \gamma0}$ are 
the density parameters of the total non-relativistic matter, 
of baryons and of radiation, 
respectively, and $h=H_0/(100{\rm km/s/Mpc})$. 

In order to compare the  
first acoustic peak in the CMB spectrum predicted 
by our LTB universe model
to the observational results, it is convenient to use the quantity 
$\mathcal S$ defined by~\cite{Alnes:2005rw}
\begin{equation}
\mathcal S:=\frac{\ell_1}{\ell_1^{\rm WMAP}}\,, 
\end{equation} 
where $\ell_1$ is the first acoustic peak in the CMB spectrum and 
$\ell_1^{\rm WMAP}=220.8$ is the mean value of the first peak
in the WMAP data~\cite{Hinshaw:2006ia} (see Table~\ref{tab:2ndand3rd}). 
Since the 1$\sigma$ error in the WMAP data is $\Delta\ell_1=0.7$, 
models with $|\mathcal S-1|>0.0063$ are ruled out at 67\% CL.  

Before closing this subsection, it may be worthwhile
to note the role of the instantaneous decoupling approximation.
Its role is to identify a homogeneous universe model which
is a good approximation to our universe model 
around the time of decoupling. Once the parameters of
a homogeneous universe model are identified, 
the positions of the acoustic peaks
are calculated from the fitting formula (30) derived
from numerical simulations for homogeneous universe models. 
Hence except for the effect of errors in the estimation 
of the cosmological parameters, there will be no
deviation of the acoustic peaks due to
the instantaneous decoupling approximation.

\subsection{Position of the first acoustic peak in the LTB model}

Because the inhomogeneity in our LTB model consists of only  
growing modes, the inhomogeneity decreases as we go back 
to the early stage. Actually, our LTB model can be well 
approximated by the Einstein-de Sitter (EdS) universe
at the decoupling time. 
In this subsection, we therefore approximate our LTB model
at the decoupling epoch by an EdS universe model.
Hereafter, we place a bar over a quantity in this EdS universe.
We determine it by setting the decoupling temperature $\bar T_D$
and the baryon-to-photon ratio $\bar\eta$ in 
this EdS universe model equal to $T_*$ on the LSS 
in our LTB model and $\eta_*$ given by Eq.~(\ref{eq:eta-star}),
respectively.
Further, we set the present CMB temperature 
to be equal to $T_0$ also in the EdS model.
These three conditions determine $\bar{\omega}_{\gamma}$,
$\bar{\omega}_{\rm b}$ and $\bar{\omega}_{\rm m}$,
and the Hubble parameter $\bar{H}_{0}$ at present,
\begin{equation}
\bar{H}_{0}^2=H_*^2\left(\frac{T_0}{T_*}\right)^3
=\frac{8\pi G}{3}\rho_*\left(\frac{T_0}{T_*}\right)^3. 
\end{equation}

Using the definition of $\alpha$, 
we have 
\begin{equation}
\bar{H}_{0}^2=\frac{8\pi G}{3}\rho_0\alpha_*
=[1+\tilde k(0)]H_0^2\alpha_*\simeq0.09120H_0^2\alpha_*\,.
\label{eq:Hbar0}
\end{equation}
Since the present CMB temperature is $T_0$, we have 
\begin{equation}
\bar{\omega}_{\gamma}
=4.2\times 10^{-5}\left(\frac{T_0}{2.725{\rm K}}\right)^4. 
\end{equation}
The quantity $\bar{\omega}_{\rm b}$ is 
expressed in terms of 
the baryon-to-photon ratio as 
\begin{eqnarray}
\bar{\omega}_{\rm b}
=\frac{8\pi G}{3\bar{H}_0^2}\eta_*\bar{n}_{\gamma0}m_{\rm p}\bar{h}^2
=3.658\times 10^7\eta_*\left(\frac{T_0}{2.725{\rm K}}\right)^3, 
\end{eqnarray}
where $m_{\rm p}$ is the proton mass. 
As for the matter density, we have $\bar\Omega_{\rm m0}=1$
by assumption, hence 
\begin{equation}
\bar{\omega}_{\rm m}=\bar\Omega_{\rm m0}\bar{h}^2=
\bar{h}^2=0.09120\alpha_* h^2, 
\end{equation}
where we have used Eq.~(\ref{eq:Hbar0}).

By using Eqs.~(\ref{eq:eta-star}) and (\ref{eq:alpha-result}), 
we can evaluate $\bar{\omega}_\gamma$, 
$\bar{\omega}_{\rm b}$ and $\bar{\omega}_{\rm m}$. 
Then, substituting $\omega_{\gamma}=\bar{\omega}_\gamma$, 
$\omega_{\rm b}=\bar{\omega}_{\rm b}$
and $\omega_{\rm m}=\bar{\omega}_{\rm m}$ for Eqs.~(\ref{eq:A1}), (\ref{eq:A2}), 
(\ref{eq:A3}) and (\ref{eq:A4}), 
we find
\begin{equation}
|\bar{\mathcal S}-1|\simeq 0.075>0.0063, 
\end{equation}
where we have used Eq.~(\ref{eq:T-result}) and set $h=0.71$ and $T_0=2.725$K. 
This result implies that our LTB model cannot explain the observed
first peak position in the CMB spectrum.

\section{Matching peaks in the CMB spectrum}
\label{sec:inverseproblem}

Let us consider if we can resolve the inconsistency of our LTB model
with the observed first acoustic peak position in the CMB spectrum.
Since SNe observations of the distance-redshift relation are 
still restricted to relatively low redshifts $z<2$, 
the distance-redshift relation may not necessarily agree 
with that in the concordance $\Lambda$CDM model at $z>2$. 
Therefore we may modify the curvature function $\tilde{k}(\tilde{r})$ 
in the domain of large $r$ so that 
the position of the first peak in an LTB model agrees 
with the observed one without affecting 
the distance-redshift relation at $z<2$.

\begin{figure}[htbp]
\begin{center}
\includegraphics[scale=1.8]{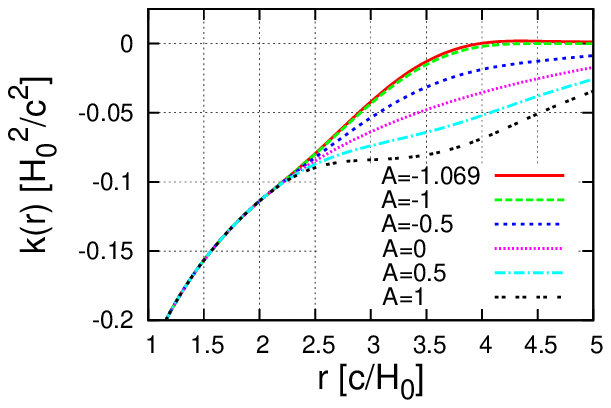}
\includegraphics[scale=1.8]{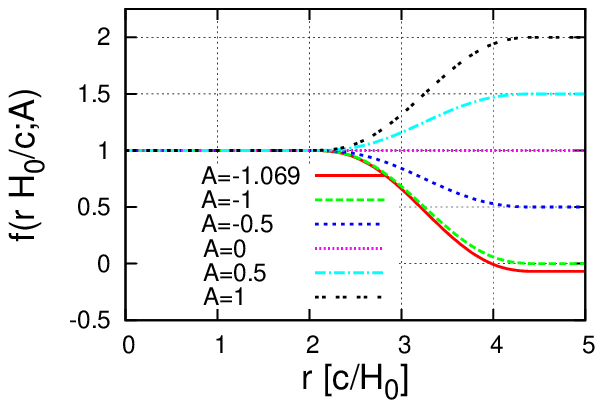}
\caption{$\tilde k(\tilde r;A)$ and $f(\tilde r;A)$ as functions
of $\tilde r$ for several values of $A$.
}
\label{fig:modk}
\end{center}
\end{figure}

As soon as we allow modifications of our LTB model at high
redshifts, we have too much freedom to fix the model uniquely.
Therefore, for simplicity, we consider a single-parameter
modification. Namely, we adopt the curvature function,
\begin{equation}
\tilde k(\tilde r;A)=\tilde k_{\rm fit}(\tilde r)\times f(\tilde r;A),
\end{equation}
where the function $f(\tilde r;A)$ is defined by 
\begin{equation}
f(\tilde r;A)=
\left\{
\begin{array}{lll}
1~~&{\rm for}~~\tilde r<2\,,\\
\displaystyle
1+\frac{16A\left(\tilde r-2\right)^3
\left(323-123\tilde r+12\tilde r^2\right)}{3125}
~~&{\rm for}~~2\leq \tilde r<9/2\,,\\
1+A~~&{\rm for}~~9/2\leq \tilde r\,.
\end{array}
\right.
\end{equation}
The functions $\tilde{k}(\tilde{r};A)$ and 
$f(\tilde r;A)$ 
are depicted in Fig.~\ref{fig:modk} for several values of $A$. 

\begin{figure}[htbp]
\begin{center}
\includegraphics[scale=1.2]{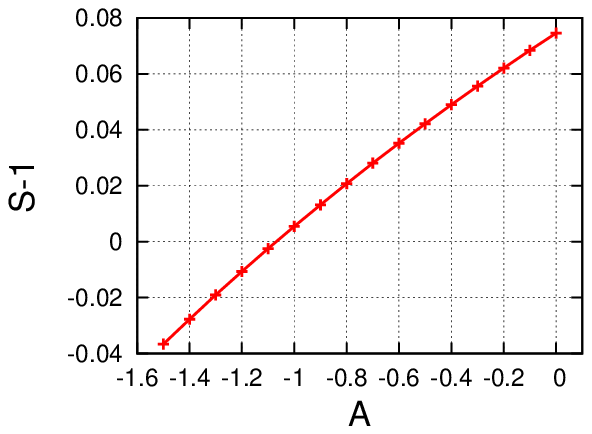}
\includegraphics[scale=1.2]{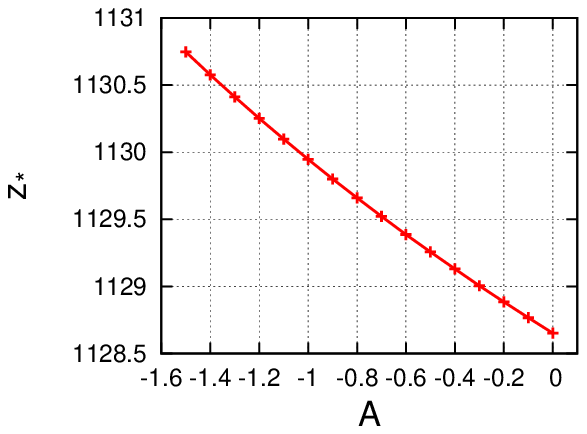}
\includegraphics[scale=1.2]{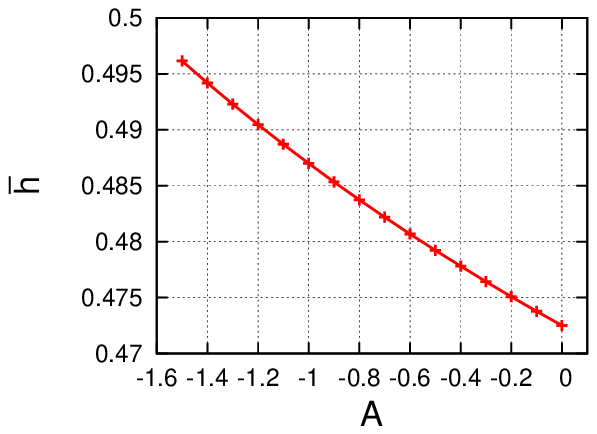}
\caption{$\mathcal S-1$(left), $z_*$(center) and
$\bar h$(right) as a function of $A$.
}
\label{fig:amptoS}
\end{center}
\end{figure}

\begin{figure}[htbp]
\begin{center}
\includegraphics[scale=1.8]{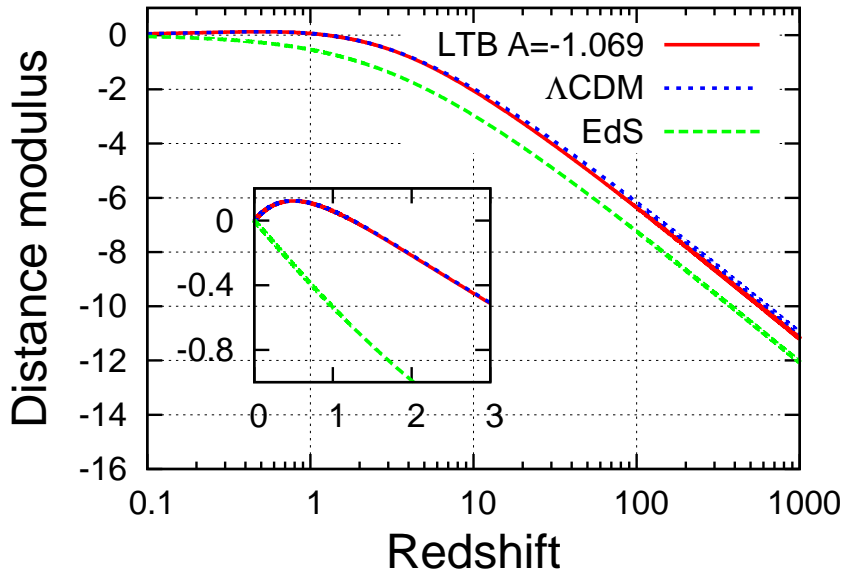}
\caption{The distance modulus for the LTB universe model 
with $t_{\rm B}=0$, 
$\tilde k(\tilde r)=\tilde k_{\rm fit}(\tilde r)f(\tilde r;-1.069)$. 
}
\label{fig:dmodulus}
\end{center}
\end{figure}

As in the previous section, we adopt the value 
$\eta_*=6.2\times10^{-10}$. Then we can evaluate 
$\mathcal S-1$ for each value of $A$. 
$\mathcal S-1$ is depicted as a function of $A$ in Fig.\ref{fig:amptoS}. 
%
%
As shown in this figure, 
$\mathcal S$ almost vanishes at $A=-1.069$. 
This fact means that the modified LTB model with $t_{\rm B}=0$, 
$\tilde k=\tilde k(\tilde r;-1.069)$,
$\eta_*=6.2\times10^{-10}$ and $h=0.71$ 
is consistent with the observed distance-redshift relation and 
the position of the first acoustic peak in 
the CMB spectrum simultaneously. 
Then we may check if the positions of 
the second and third peaks are also consistent. 
As shown in the Table \ref{tab:2ndand3rd}, 
they indeed turn out to be consistent with the 
WMAP data~\cite{Hinshaw:2006ia}. 
%
\begin{table}[htbp]

\caption{second and third peak positions in WMAP data and our LTB model 
with $A=-1.069$. 
We have assumed $n_{\rm s}=0.963$.  }
\label{tab:2ndand3rd}
\begin{center}
\begin{tabular}{|c||c|c|c|}
\hline
~&1st&2nd&3rd\\
\hline
\hline
WMAP&$220.8\pm0.7$&$530.9\pm3.8$&700-1000\\
\hline
Our model
&220.8&529.3&782.5\\
\hline
\end{tabular}
\end{center}
\end{table}
The distance modulus in this LTB universe model with $A=-1.069$ is shown 
in Fig.~\ref{fig:dmodulus}. 
%

The resultant LTB universe model is approximated by an EdS universe 
model with the Hubble parameter $h\simeq0.49$ in the spatially
asymptotic region. This result is consistent with previous
work in the literature~\cite{Blanchard:2003du,Alnes:2005rw,
Alexander:2007xx,GarciaBellido:2008nz} 
in which the possibility for lower Hubble universes to 
explain the CMB spectrum without dark energy has been pointed out.


\section{Summary and Discussion}
\label{sec:sudi}
In this paper, we investigated
if an LTB universe model can account 
not only for the observed distance-redshift relation 
but also for the observed peak positions in the CMB spectrum,
assuming that we are at the symmetry center of the universe.

First, we presented an LTB model 
whose distance-redshift relation is equal to
that of the concordance $\Lambda$CDM model in the whole redshift domain.
This model is expressed in terms of a fitting function for the curvature 
numerically obtained in Ref.~\cite{Yoo:2008su}. 
Then, we determined the last scattering surface for photons
in this model by numerically integrating the past-directed
radial null geodesics emanating from the central observer 
until Gamow's criterion is satisfied.

The metric of this LTB model can be well approximated by the 
EdS universe at the decoupling epoch.
Thus, assuming that our model universe experiences the same
history as that of a homogeneous and isotropic universe
before decoupling, we evaluated the positions of the acoustic peaks 
in the CMB spectrum.
It was found that thus obtained position of the first acoustic peak 
deviates from the observed position more than 7\%,
implying that our LTB universe model whose distance-redshift relation 
agrees with that of the concordance $\Lambda$CDM model 
for the entire redshift domain is ruled out. 

In order to resolve this inconsistency, 
we considered modifications of the LTB model in the region 
far from the symmetry center, corresponding to the redshift
domain of $z>2$. Since the observation of Type 
Ia supernovae are limited within $z<2$, 
this modification does not contradict the current observational data. 
Then we considered a one-parameter family of LTB models
whose distance-redshift relation matches the observational data
at $z<2$, and found a parameter that gives
the first acoustic peak position consistent with
the observed peak position.
Then we checked the second and third peak positions 
and found that they are also consistent with the 
WMAP data~\cite{Hinshaw:2006ia}. 

We have not discussed relative heights of
acoustic peaks in this paper. At the moment, we have no
idea if one can also fit them by an LTB model.
There is, however, a hint in the literature.
In Ref.~\cite{Blanchard:2003du}, it was reported that 
the observed CMB spectrum can be regarded as that 
in the EdS universe with a low value of the
Hubble constant if one tunes the primordial power spectrum 
of the density perturbation. If this is the case,
then we may be able to find an LTB model
which can account for the observed relative heights of 
the acoustic peaks. This issue is left for future work.

In this paper, we have not introduced 
isocurvature components in the inhomogeneity 
on the last scattering surface. 
In a forthcoming paper, we plan to investigate
if we may construct an LTB model consistent with
the observation of the kinematic Sunyaev-Zel'dovich effect. 
Our preliminary analysis indicates that
isocurvature perturbations between the non-relativistic matter
and photons are necessary in order to explain it.
However, if we introduce large radial isocurvature perturbations 
on the decoupling hypersurface, the history of the inhomogeneous 
universe model before decoupling would differ substantially
from the conventional adiabatic perturbation scenario in the
homogeneous and isotropic background.
In such a case, our analysis of the acoustic peaks 
in the CMB spectrum may be invalidated.
However, since we have considered only a single-parameter
family of LTB models, while there still remains a
functional degree of freedom, it is premature to
make any definite statement at the moment. 
Apparently much more work seems necessary to
exclude or select the LTB universe as a viable
alternative model of our universe.

\begin{acknowledgments}
C.M. would like to thank the organizers and all the participants 
of the workshop LLTB2009 in KEK Tsukuba, Japan, for helpful comments. 
C.M acknowledges the
Korea Ministry of Education, Science and Technology (MEST) for the support
of the Young Scientist Training Program at the Asia Pacific Center for
Theoretical Physics (APCTP). 
This work was also supported in part by 
Korea Institute for Advanced Study under the KIAS Scholar program,  
by the Grant-in-Aid for the Global COE Program 
``The Next Generation of Physics, Spun from Universality and Emergence''
from the Ministry of Education, Culture, 
Sports, Science and Technology (MEXT) of Japan, 
by Grant-in-Aid for Scientific Research (C), No.~21540276 
from the Ministry of Education, Science, Sports and Culture,
by JSPS Grant-in-Aid for Scientific Research (A) No.~21244033,
and by JSPS Grant-in-Aid for Creative Scientific Research No.~19GS0219.
\end{acknowledgments}

\appendix

\section{peak positions in homogeneous universes}
\label{sec:peakposihome}

\subsection{expression for the sound horizon}

Since the universe considered in this paper is well 
approximated by the homogeneous and isotropic universe with vanishing 
spatial curvature 
near the decoupling epoch, we consider the only 
such a universe model here. The infinitesimal world interval is given by 
\begin{equation}
ds^2=-dt^2+a^2(t)(dr^2+r^2d\Omega^2).
\end{equation}
The scale factor $a(t)$ is normalized so that its present value $a_0$ is unity.
 
The sound horizon $r_{\rm s}$ is defined by 
\begin{equation}
r_{\rm s}=\int^t_0\frac{c_s}{a}dt=\int^a_0\frac{c_s}{a^2H}da, 
\label{eq:sh1}
\end{equation}
where $c_{\rm s}$ is the sound speed given by 
\begin{eqnarray}
c_{\rm s}&=&\frac{c}{\sqrt{3(1+R)}},
\label{eq:sounds}\\
R&:=&\frac{3\rho_b}{4\rho_\gamma}
=\frac{3\omega_{\rm b}}{4\omega_\gamma}\frac{T_0}{T}
=\frac{3\omega_{\rm b}}{4\omega_\gamma}\frac{1}{1+z}. 
\end{eqnarray}
At the matter-radiation equality time, we have 
\begin{equation}
R_{\rm eq}:=
\frac{3\omega_{\rm b}}{4\omega_\gamma}\frac{T_0}{T_{\rm eq}}
=\frac{3\omega_{\rm b}}{4\omega_{\rm m}}, 
\end{equation}
where we have used that the temperature $T_{\rm eq}$ at the equality time 
can be written as 
\begin{equation}
T_{\rm eq}=\frac{\omega_{\rm m}}{\omega_{\gamma}}T_0. 
\end{equation}
The denominator of the integrand in (\ref{eq:sh1}) can be written as 
\begin{equation}
a^2H
=\sqrt{\Omega_{\rm m0}a_{\rm eq}
\left(\frac{R}{R_{\rm eq}}+1\right)},
\label{eq:a2h}
\end{equation}
where $a_{\rm eq}$ is the scale factor at the equality time, $\Omega_{\rm m0}$ is 
the density parameter of the total non-relativistic matter 
and we have used the relation $R=aR_{\rm eq}/a_{\rm eq}$.
Using the expressions (\ref{eq:sounds}) and (\ref{eq:a2h}), 
we can perform the integration in (\ref{eq:sh1}) as
\begin{equation}
r_{\rm s*}=\frac{2}{k_{\rm eq}}\sqrt{\frac{2}{3R_{\rm eq}}}
\ln\left(\frac{\sqrt{1+R_*}
+\sqrt{R_*+R_{\rm eq}}}{1+\sqrt{R_{\rm eq}}}\right), 
\end{equation}
where 
\begin{equation}
k_{\rm eq}=\frac{H_0}{c}\sqrt{\frac{2\Omega_{\rm m0}}{a_{\rm eq}}}
=\frac{H_{\rm 0}}{c}\sqrt{2\Omega_{\rm m0}\frac{T_{\rm eq}}{T_0}}
=4.714\times10^{-4}\frac{\omega_{\rm m}}{\sqrt{\omega_\gamma}}h^{1/2}
{\rm Mpc^{-1}}\,,
\end{equation}
and
\begin{equation}
R_*=\frac{3\omega_{\rm b}}{4\omega_\gamma}\frac{1}{1+z_*}\,. 
\end{equation}

\subsection{small shift parameters for the first, second and third peaks}

We define the ratio $r_\star$ of the radiation density to the matter 
density at decoupling as 
\begin{equation}
r_\star=\frac{\rho_{\gamma *}}{\rho_{{\rm m }*}}
=\frac{\omega_{\gamma }}
{\omega_{{\rm m }}}\frac{T_*}{T_0}
=\frac{\omega_{\gamma }}
{\omega_{{\rm m }}}(1+z_*)\,. 
\label{eq:A1}
\end{equation}
Using this ratio, a fitting formula given in Ref.~\cite{Doran:2001yw}
gives 
\begin{eqnarray}
\phi_1&=&a_1r_\star^{a_2}, \label{eq:A2}\\
\phi_2&=&\phi_1+\delta\phi_2=
\phi_1+c_0-c_1r_\star-c_2r_\star^{-c_3}+0.05(n_{\rm s}-1), \\
\phi_3&=&e_1r_\star^{e_2}-0.037(n_{\rm s}-1), 
\end{eqnarray}
where 
\begin{eqnarray}
a_1&=&0.286+0.626\omega_{\rm b}, \label{eq:A3}\\
a_2&=&0.1786-6.308\omega_{\rm b}+174.9\omega_{\rm b}^2-1168\omega_{\rm b}, \label{eq:A4}\\
c_0&=&-0.1+0.213\exp(-52\omega_{\rm b}), \\
c_1&=&0.063\exp(-3500\omega_{\rm b}^2)+0.015, \\
c_2&=&6\times 10^{-6}+0.137(\omega_{\rm b}-0.07)^2, \\
c_3&=&0.8+70\omega_{\rm b}, \\
e_1&=&0.302-2.112\omega_{\rm b}+0.15\exp(-384\omega_{\rm b}), \\
e_2&=&-0.04-4.5\omega_{\rm b}. 
\end{eqnarray}


\end{document}